\documentclass[aps,showpacs,showkeys,amsmath,amssymb,preprint,prd,%
floatfix,superscriptaddress]{revtex4}

\usepackage{graphicx}
\usepackage{amssymb}
\usepackage{dcolumn}% Align table columns on decimal point
\usepackage{docs}%
\usepackage{bm}% bold math
\begin{document}
\title{Shear and bulk viscosities of the gluon plasma in a quasiparticle 
description}

\author{M.~Bluhm}
\affiliation{SUBATECH, UMR 6457, Universit\'{e} de Nantes, 
Ecole des Mines de Nantes, IN2P3/CNRS. 4 rue Alfred Kastler, 
44307 Nantes cedex 3, France}
% address was wrong CRNS --> CNRS
\affiliation{CERN, Physics Department, Theory Devision, 
CH-1211 Geneva 23, Switzerland}
\author{B.~K\"ampfer}
\affiliation{Institut f\"ur Strahlenphysik, 
Helmholtzzentrum Dresden-Rossendorf, PF 510119, 01314 Dresden, Germany}
\affiliation{Institut f\"ur Theoretische Physik, TU Dresden, 
01062 Dresden, Germany}
\author{K.~Redlich}
\affiliation{CERN, Physics Department, Theory Devision, 
CH-1211 Geneva 23, Switzerland}
\affiliation{Institute of Theoretical Physics, University of Wroclaw, 
PL-50204 Wroclaw, Poland}

%\date{\today}

%Use showkeys class option if keyword display desired
%PACS, the Physics and Astronomy Classification Scheme.

\begin{abstract}
% last edited on 16/11/2010
Shear and bulk viscosities of deconfined gluonic matter are investigated 
within an effective kinetic theory by describing the strongly 
interacting medium phenomenologically in terms of quasiparticle excitations 
with medium-dependent self-energies. We show that the resulting 
transport coefficients reproduce the parametric dependencies on temperature 
and coupling obtained in perturbative QCD at large temperatures and small 
running coupling. The extrapolation into the non-perturbative regime results 
in a decreasing specific shear viscosity with decreasing temperature, 
exhibiting a minimum in the vicinity of the deconfinement transition, 
while the specific bulk viscosity is sizeable in this region falling off rapidly 
with increasing temperature. The temperature dependence of specific shear and bulk 
viscosities found within this quasiparticle description of the pure gluon 
plasma is in agreement with available lattice QCD results. 
\end{abstract}
%%%%%%%%%%%%%%%%%%%%%%%%%%%%%%%%%%%%%%%%%%%%%%%%%%%%%%%%%%%%%%%%%%%%%%%%%%%%%%

\maketitle

\section{Introduction \label{sec:1}}
% last edited on 08/11/2010

Transport properties of strongly interacting matter, encoded in 
coefficients such as shear and bulk viscosities, which describe 
the hydrodynamic response of the system to energy and momentum 
density fluctuations, are of particular importance for understanding 
the nature of QCD matter. Their firm knowledge, besides other 
characteristics like the QCD equation of state (EoS), has great impact 
on a variety of physical phenomena in cosmology, astrophysics or 
nuclear physics. In particular, the dynamical description of the 
medium created in relativistic heavy-ion collision experiments 
requires a precise determination of the temperature $T$ and density 
dependencies of these transport coefficients. Even though tremendous 
efforts have been invested both from theoretical as well as 
experimental sides, this remains still a challenging task, especially 
in the regime of experimentally accessible temperatures and densities. 

It was shown that the observed collective flow behaviour of the matter 
created at the Relativistic Heavy Ion Collider (RHIC), in particular 
the measured elliptic flow~\cite{Adams05,Adcox05,Arsene05,Back05}, 
could quantitatively be described by means of ideal 
hydrodynamics~\cite{Kolb01,Huovinen01,Teaney01,Hirano02,Kolb03,Kolb04} for 
not too large transverse momenta of the particles. 
Although the results of these calculations depend crucially on the 
employed freeze-out prescription as well as the assumed initial 
conditions~\cite{Miller03,Hirano06,Luzum08,Petersen09,Hama09,
Werner09,Song09,Hirano09}, the success of ideal 
hydrodynamics suggested at most small dissipative 
effects in the medium produced at RHIC~\cite{Heinz02,Lacey07,Drescher07}. 
This observation led to the interpretation that the produced quark-gluon plasma 
(QGP) behaves as a strongly coupled fluid rather than a weakly interacting 
gas~\cite{Gyulassy05,Shuryak05,Heinz05}. Detailed calculations within 
the framework of relativistic dissipative fluid dynamics were 
performed~\cite{Muronga02,Teaney03,Muronga04,Heinz06,Baier06,Chaudhuri06,Romatschke07,
Chaudhuri0708,Dusling08,Luzum08,Molnar08,Muronga08,Song08,Fries08,Song10}, confirming 
indeed the smallness of dissipative effects. Therein, the equations of (for many EoS) 
causal Israel-Stewart theory~\cite{Israel79,Hiscock83} are solved. Recently, 
however, corrections to this second-order in gradients theory were 
reported~\cite{Baier08,Betz08,York08}. 

On the other hand, some transport coefficients, in particular the shear viscosity 
$\eta$, cannot be arbitrarily small. In fact, the quantum mechanical uncertainty 
relation sets a fundamental lower bound on $\eta$ for any physical 
system~\cite{Danielewicz85}. Based on unitarity 
arguments~\cite{Policastro01,Kovtun03,Kovtun05}, the dimensionless 
specific shear viscosity (i.~e.~the ratio of shear viscosity to entropy density $s$) 
was recently conjectured to be bound from below by 
$(\eta/s)_{KSS}\ge \hbar/(4\pi)$ (Kovtun-Son-Starinets bound). 
The comparison of causal viscous hydrodynamic calculations with data on low-momentum 
elliptic flow spectra measured at RHIC showed that $\eta/s$ should be close to this 
lower bound~\cite{Luzum08,Romatschke07,Chaudhuri0708,Gavin06,Lacey07,Drescher07}. 
Nevertheless, some other results seem to indicate that the KSS bound might be 
violated for the QGP at RHIC~\cite{Romatschke07,Song08,Lacey07}. Moreover, there 
are other physical systems~\cite{Kats07,Cohen07,Cherman08,Jakovac10}, where 
this is indeed the case. 

The specific shear viscosity $\eta/s$ as a function of temperature is known to 
exhibit a minimum in the vicinity of a phase transition for a variety of 
liquids and gases~\cite{Kovtun05,Csernai06} as well as for ultracold fermionic 
systems close to the unitarity limit~\cite{Schaefer07}. This gave rise to the 
conjecture~\cite{Lacey07,Csernai06} that $\eta/s$ for QCD matter exhibits a similar 
behaviour in the vicinity of the deconfinement phase transformation.

In contrast, the bulk viscosity $\zeta$ is exactly zero for conformally 
invariant systems. This is due to the vanishing trace of the energy-momentum tensor 
in conformal theories. In QCD, as quantum fluctuations break the scale invariance, 
the bulk viscosity is non-zero at finite $T$ and approaches zero only for 
asymptotically large $T$. In the vicinity of the deconfinement phase 
transformation, the dimensionless specific bulk viscosity $\zeta/s$ is expected 
to become large and even to diverge at a second-order phase 
transition~\cite{Kharzeev08,Karsch08,Romatschke09,Paech06,Moore08,Sasaki09-1,Sasaki09-2}. 
However, there are examples of strongly coupled theories~\cite{Buchel08,Buchel09}, 
for which the contrary is found. 

Transport coefficients of strongly interacting systems have been calculated 
in a variety of different approaches. Despite the fact that lattice QCD studies 
of dynamical quantities, such as viscosities, are tedious to perform, they represent 
the first-principle non-perturbative method in the strong coupling regime. Recently, first 
results for shear and bulk viscosities obtained within lattice QCD simulations 
were reported in~\cite{Meyer07,Meyer08,Nakamura05,Sakai07}. 

In quantum field theories, calculations of transport coefficients are often 
performed within the framework of linear response theory resulting in the known 
Kubo formulae~\cite{Kubo57,Hosoya84,Schafer09}. This approach for determining 
shear and bulk viscosities was exploited for a weakly coupled scalar field theory 
in~\cite{Jeon95,Jeon96}. Likewise, kinetic theory can be applied as a rigorous tool 
to quantify transport coefficients at weak coupling from a linearized Boltzmann 
equation. In fact, it was shown for weakly coupled scalar fields~\cite{Jeon95,Jeon96} 
and for hot QED~\cite{Gagnon07,Gagnon07-1} that both approaches yield equivalent 
results, when a linearized Boltzmann equation for excitations with temperature dependent 
self-energies and scattering amplitudes is considered. In QCD, kinetic-theory 
weak-coupling results were reported for the shear viscosity at leading logarithmic 
order~\cite{Arnold00} and at full leading order~\cite{Arnold03} in the QCD running 
coupling $\alpha_s$. These calculations found for $\alpha_s\lesssim 0.25$ a specific 
shear viscosity $\eta/s > 1$. For the bulk viscosity perturbative QCD (pQCD) leading 
order results were presented in~\cite{Arnold06}. 

Further investigations of transport coefficients utilize e.~g.~a specific ansatz 
for the spectral function entering the Kubo-expression for 
$\eta$~\cite{Peshier05,Aarts02,Aarts04} or apply sum rules for the spectral density 
of the trace of the energy-momentum tensor, thereby relating $\zeta$ with basic 
thermodynamic quantities~\cite{Kharzeev08,Karsch08,Romatschke09}. Moreover, within 
a parton cascade model based on kinetic theory~\cite{Xu05} a value for 
$\eta/s\sim 0.5$ was found~\cite{Xu08}, which is much smaller than the pQCD result 
$\eta/s\sim 2.7$~\cite{Arnold00,Arnold03} at comparable coupling $\alpha_s\sim 0.1$. 
Recent similar studies report $\eta/s\sim 0.8$~\cite{82A} and 
$\eta/s\sim 1$~\cite{Chen09,83A} instead. 

In this work, we study the properties of shear and bulk viscosities for the pure 
gluon plasma at finite temperature. The dynamics of the system is considered 
phenomenologically by assuming that it can be described in terms of gluonic 
quasiparticle excitations. In the underlying quasiparticle model (QPM), the 
(quasi)gluon excitations obey a medium-dependent dispersion relation, 
where the corresponding self-energy depends on $T$ both explicitly but also 
implicitly via a phenomenological effective coupling 
$G^2(T)$~\cite{Peshier94,Peshier96,Peshier00,Peshier02}. In a pure gluon plasma, 
there are no conserved charges related to internal symmetries. Thus, $T$ is the 
only thermal variable characterizing properties of the system. The QPM was 
successfully employed to quantitatively describe lattice QCD results of 
equilibrium thermodynamics such as the EoS and related 
quantities~\cite{Bluhm05,Bluhm07-1,Bluhm07-2,Bluhm08-1,Bluhm08-2}. In the 
following, we apply this model to systems exhibiting small departures 
from thermal equilibrium by using the results of an effective kinetic 
theory approach~\cite{Jeon95,Jeon96} for shear and bulk viscosities derived 
for a quasiparticle system in~\cite{Chakraborty10}. 

We show that this quasiparticle picture combined with kinetic theory 
provides a powerful method to quantify the transport properties 
of the gluon plasma in a broad 
temperature range above the critical temperature for deconfinement $T_c$. In 
particular, we show that in the high temperature limit the resulting transport 
coefficients exhibit parametrically the dependencies on $T$ and the coupling known from 
pQCD. We also show that our results extrapolated into the non-perturbative regime 
are in agreement with available lattice QCD results.

Our paper is organized as follows. In the next section, we 
recall the first-order non-equilibrium corrections to the energy-momentum tensor 
for the (quasi)gluon plasma and discuss features of its viscosity coefficients in relaxation 
time approximation as well as specify the employed quasiparticle parametrization.
In sections~\ref{sec:eta} and \ref{sec:zeta}, the shear and bulk viscosities 
are quantified and compared with available lattice QCD data.
The summary of our results is found in section~\ref{sec:4}.  
Throughout this work we use natural units, i.~e.~$\hbar=c=k_B=1$. 

\section{Shear and bulk viscosities \label{sec:3new}}

For charge-neutral systems and in the absence of additional conservation laws, 
the local conservation equations for energy and momentum in relativistic
hydrodynamics as effective theory of long-wavelength modes 
determine the evolution of the system. Phenomenologically, the energy-momentum 
tensor $T^{\mu\nu}$, 
with equations of motion $\partial_\nu T^{\mu\nu} = 0$,
is defined as decomposition into a perfect fluid (or thermal 
equilibrium) part $T^{\mu\nu}_{(0)}$ and a part $T^{\mu\nu}_{(1)}$, 
which accounts for first-order dissipative 
effects~\cite{deGroot,Landau,LandauN}: 
$T^{\mu\nu}  =  T^{\mu\nu}_{(0)}+T^{\mu\nu}_{(1)}$ with 
$T^{\mu\nu}_{(0)}  =  \epsilon u^\mu u^\nu -P\Delta^{\mu\nu}$ and 
$T^{\mu\nu}_{(1)} =  \zeta \Delta^{\mu\nu} \partial_\alpha u^\alpha + \eta 
 S^{\mu\nu}_{\,\,\,\,\,\,\alpha\beta} \partial^\alpha u^\beta$, 
where $\epsilon$ and $P$ denote energy density and thermodynamic pressure 
of the system, respectively; both are related through the 
equation of state $P=P(\epsilon)$. The projector
$\Delta^{\mu\nu}=g^{\mu\nu}-u^\mu u^\nu$ with 
normalized four-velocity $u_\mu u^\mu=1$ and metric $g^{\mu\nu}$ is orthogonal to $u^\mu$; 
the projector $S^{\mu\nu}_{\,\,\,\,\,\,\alpha\beta}=
\Delta^\mu_\alpha \Delta^\nu_\beta + 
\Delta^\mu_\beta \Delta^\nu_\alpha - 2\Delta^{\mu\nu} \Delta_{\alpha\beta}/3$ 
is also $u^\mu$-orthogonal. 
The coefficients $\eta$ and $\zeta$ entering $T^{\mu\nu}_{(1)}$
denote shear and bulk viscosities, respectively.
Note that if 
$T^\mu_{\,\,\,\,\mu}$ can be formulated solely as a function of $\epsilon$, then 
$\zeta=0$ is found~\cite{Weinberg,Weinberg71}. 
In the way $T^{\mu\nu}$ is defined,
one takes into account that from $\partial_\nu T^{\mu\nu}=0$ only 
four out of ten independent components of $T^{\mu\nu}$ can be determined, while 
the others follow from expanding $T^{\mu\nu}$ around $T^{\mu\nu}_{(0)}$ in terms 
of small space-time gradients of the fluid four-velocity field 
$u^\mu$~\cite{Weinberg71,Weinberg}.
This expansion holds up to including further gradients or higher powers 
of $u^\mu$. 
The Landau-Lifshitz condition~\cite{deGroot,Landau}
$\epsilon\equiv T^{\mu\nu}u_\mu u_\nu=T^{\mu\nu}_{(0)}u_\mu u_\nu$
provides the only uniform definition of local 
flow for systems without additional conserved currents~\cite{Danielewicz85}. 

\subsection{Basic definitions}

Concise expressions for the shear and bulk viscosities 
have been derived systematically in~\cite{Chakraborty10}
within a quasiparticle framework. 
In relaxation time ($\tau$) approximation, one can deduce for the (quasi)gluon plasma 
\begin{eqnarray}
 \label{equ:eta2} 
 \eta &=& \frac{1}{15T} \int \frac{d^3 \vec{p}}{(2\pi)^3} 
 n(T)(1+d^{-1}n(T)) \frac{\tau}{E^2} \vec{p}^{\,4} , \\
 \label{equ:zeta3}
 \zeta &=& \frac{1}{T} \int \frac{d^3 \vec{p}}{(2\pi)^3} 
 n(T)(1+d^{-1}n(T)) \frac{\tau}{E^2} 
 \left\{ \left[E^2-a\right]\frac{\partial P}{\partial\epsilon} 
 - \frac13 \vec{p}^{\,2} \right\}^2 ,
\end{eqnarray}
where $n(T)=d(e^{E/T}-1)^{-1}$ with $d = 16$ is the equilibrium distribution function, 
$\partial P/\partial\epsilon$ denotes the squared speed of sound 
and $a \equiv T^2 \partial \Pi(T) / \partial T^2$. The excitations
obey the dispersion relation $E^2 = {\vec p\,}^2 + \Pi (T)$ with $\Pi (T)$ as
temperature dependent self-energy. The equilibrium distribution $n(T)$ is a 
particular solution
of an effective kinetic theory of Boltzmann-Vlasov type~\cite{deGroot} with
force 
$F^\alpha (x) = (\partial^\alpha \Pi (x))/(2 \sqrt{\Pi (x)})$~\cite{Jeon95,Jeon96}, 
satisfying $p_\alpha F^\alpha = 0$ for the momentum 
$p^\alpha \sim (E(\vec p, x), \vec p \,)$, in a state without entropy production
where the collision term vanishes. Thermodynamic self-consistency is
ensured, i.e.\   
$\epsilon(T) \equiv T^{\mu\nu}_{(0)} u_\mu u_\nu$,
$P(T) \equiv -\frac13 T^{\mu\nu}_{(0)} \Delta_{\mu\nu}$
and $\epsilon = T^2 \partial (P(T)/T)/\partial T$
from $\epsilon + P = T s$ with $s = \partial P / \partial T$. 

Equation (\ref{equ:eta2}) for the shear viscosity, which was first derived
in the quasiparticle framework in~\cite{Bluhm09}, 
coincides formally with the results~\cite{Hosoya85,Gavin85} obtained within kinetic theory
without the force term $\propto F^\alpha$ and without self-energy. 
Nonetheless, it differs from these results, 
as the effective quasiparticle mass $\Pi(T)$ entering $E$ 
reflects medium effects. This was also observed 
in~\cite{Sasaki09-1,Sasaki09-2,Toneev10,Toneev10new}.

The expression (\ref{equ:zeta3}) for the bulk viscosity 
represents a special case of the general form of $\zeta$,
$\frac{1}{T} \int \frac{d^3 \vec{p}}{(2\pi)^3} 
 n(T)(1+d^{-1}n(T)) \frac{\tau}{E^2} 
 \left\{ \left[E^2-a\right]\frac{\partial P}{\partial\epsilon} 
 - \frac13 \vec p^{\, 2} \right\}
 \left\{X [E^2 - a] - \frac13 \vec{p}^{\, 2} \right\}$,
obtained by setting the momentum independent factor 
$X = \partial P / \partial \epsilon$ for invoking the Landau-Lifshitz condition.
(Choosing instead $X = \frac13$,
$\frac{1}{T} \int \frac{d^3 \vec{p}}{(2\pi)^3} 
 n(T)(1+d^{-1}n(T)) \frac{\tau}{E^2} \frac13
 \left\{ \left[E^2-a\right]\frac{\partial P}{\partial\epsilon} 
 - \frac13 \vec p^{\, 2} \right\}
 \left\{ \Pi (T) - a \right\}$
would follow for $\zeta$, which corrects a factor-2 mistake in the result 
derived in~\cite{Bluhm09}.) 
Equation~(\ref{equ:zeta3}) differs even formally from the previous 
results~\cite{Hosoya85,Gavin85,Toneev10,Sasaki09-1,Sasaki09-2,Toneev10new}. 
This is because it contains the subtracted mass term $(\Pi(T)-a)$ as 
in~\cite{Jeon96,Arnold06,Bluhm09} and thus 
also the derivative of $G^2(T)$ with respect to $T$ and not only 
$\Pi(T)$ alone. If one would use a temperature independent mass squared $M^2$ 
instead of $\Pi(T)$ in $E$, then both results for $\zeta$ would correspond 
to the expression obtained in~\cite{Gavin85}. 

Medium (i.~e.~EoS) effects influence the bulk viscosity more prominently than 
the shear viscosity. This is because $\zeta$, just opposite to $\eta$, 
is relevant in processes, which change the volume of a system rather than 
its shape. Then, in the pure gluon plasma, the only way to relax locally the 
system to equilibrium is by changing the average energy of the excitations 
and, thus, the energy density. 

Discussing the general properties of $\zeta$ from Eq.~(\ref{equ:zeta3}), we note that 
for a gas of particles with small and constant, i.~e.\ temperature independent, 
mass a bulk viscosity $\zeta\simeq 0$ is found in the ultra-relativistic limit 
($p^\alpha p_\alpha \simeq 0$) because $\partial P/\partial\epsilon\simeq 1/3$. 
In the non-relativistic case, $T_{(1)\,i}^{\,\,\,\,i} \equiv 2 \, T_{(1)}^{00}$ 
is obtained from the above decomposition of the energy-momentum tensor, 
which in the local fluid rest frame 
vanishes due to the Landau-Lifshitz condition implying also $\zeta =0$. 
For massless particles one has exactly 
$\partial P/\partial\epsilon=1/3$, $\partial\Pi/\partial T=0$ and 
$E = |\vec{p}\,|$, such that $\zeta=0$ is found from Eq.~(\ref{equ:zeta3}) in 
line with~\cite{LandauN,Weinberg71,Hosoya85,Horsley87,Gavin85}. 

In the conformal limit, i.~e.~at asymptotically large $T$, the bulk viscosity 
vanishes, $\zeta\rightarrow 0$. This is because the squared factor in 
Eq.~(\ref{equ:zeta3}) can be rewritten as 
\begin{eqnarray}
\nonumber
 \left[E^2-a\right]\frac{\partial P}{\partial\epsilon} 
 - \frac13 \vec{p}^{\,2} & = & \vec{p}^{\,2} \left( \frac{\partial P}{\partial\epsilon} 
 - \frac13 \right) + \frac{\partial P}{\partial\epsilon} 
 \left(\Pi(T)-a\right) \\
\label{equ:charge1}
 & = & \vec{p}^{\,2} \left( \frac{\partial P}{\partial\epsilon} 
 - \frac13 \right) - \frac12 \frac{\partial P}{\partial\epsilon} 
 T^4 \frac{\partial G^2}{\partial T^2}
\end{eqnarray}
for $\Pi(T)=T^2 G^2(T)/2$ (cf.\ section \ref{sec:IIC}). 
Consequently, in the combination $(\Pi(T) -a)$ 
the leading temperature dependence of $\Pi(T)$ is canceled out, 
cf.~also~\cite{Arnold06}, leaving solely the derivative of the effective coupling 
with respect to $T$. In the conformal limit, however, the only dimensionful 
scale is the temperature. This prohibits, in particular, a running of the 
coupling with $T$, i.~e.~$\partial G^2/\partial T = 0$. 
Therefore, together with $\partial P/\partial\epsilon=1/3$ 
(cf.~Eq.~(\ref{equ:soundvelo}) below), one finds 
\begin{equation}
\nonumber
 \vec{p}^{\,2} \left( \frac{\partial P}{\partial\epsilon} 
 - \frac13 \right) - \frac12 \frac{\partial P}{\partial\epsilon} 
 T^4 \frac{\partial G^2}{\partial T^2} = 0 \,.
\end{equation}
In other words, the terms $(\partial P/\partial\epsilon - 1/3)$ and $(\Pi(T) -a)$ 
are a measure for the deviation of the system from conformal invariance. 

\subsection{Relaxation times}

To quantify shear and bulk viscosities from Eqs.~(\ref{equ:eta2}) 
and~(\ref{equ:zeta3}), respectively, one needs to specify $\tau$. 
Microscopically considered, the shear viscosity is dominantly 
determined by elastic two-body scatterings among excitations at typical 
momenta of order $\mathcal{O}(T)$. The bulk viscosity, in contrast, measures 
relaxation of disturbances, which are caused by a slow uniform expansion of 
the system. Consequently, it is determined by inelastic particle number 
changing processes and is sensitive to excitations with soft 
momenta~\cite{Jeon95,Jeon96,Arnold00,Arnold03,Arnold06,LandauN}. 
This implies that the relaxation times for $\eta$ and $\zeta$ 
are in general different.
In the following, however, we concentrate on elastic gluon-gluon 
$2\leftrightarrow 2$ scatterings only. The corresponding relaxation rate, 
$\tau^{-1}=\tilde{n}\langle \tilde{v}\sigma\rangle$, is determined from the 
thermal-averaged total (elastic) scattering cross section 
$\langle \tilde{v}\sigma\rangle$ and the particle density in local 
equilibrium $\tilde{n}$, where $\tilde{v}$ is the relative velocity between 
the two scattering particles. 

The expression we employ as ansatz for $\tau^{-1}$ originates 
from solving the Boltzmann equation by a variational treatment including 
screening effects in the gluon-gluon scattering~\cite{Heiselberg94}. The 
result depends crucially on the ratio of maximum to minimum momentum 
transfer in the scattering process. The minimum momentum transfer 
is related to the Debye screening mass $m_D$. 
The maximum momentum transfer $p_{max}$ is instead limited by the typical 
particle momentum in the medium. For our quasiparticles this is 
$4>p_{max}/T>3$ in the temperature interval $1<T/T_c<4$, 
while for larger $T$, $p_{max}/T$ decreases extremely slowly with $T$ and 
can be set constant. 

At large temperatures, the relaxation rate $\tau^{-1}$ was found to be 
proportional to $\ln (1/m_D^2)$~\cite{Heiselberg94}. In order to extend 
this perturbative QCD result into the non-perturbative regime, we replace 
the running coupling $g^2=4\pi\alpha_s$ entering $m_D$ by the effective 
coupling $G^2(T)$ from Eq.~(\ref{equ:G2QPM}) below. In this way, 
one arrives at the following phenomenological form for the inverse 
relaxation time 
\begin{equation}
\label{equ:tau}
 \tau^{-1} = a_1 T G^4 \ln (a_2/G^2) \,. 
\end{equation}
The above expression is momentum independent and is assumed to be common for 
shear and bulk viscosities. This may be considered as a rather crude 
approximation~\cite{Arnold00,Arnold06,Heinz94}. A more 
refined approach, employing a momentum dependent relaxation time 
including elastic and inelastic scatterings, was recently 
proposed in~\cite{Chakraborty10}. 

\subsection{Temperature dependent self-energy \label{sec:IIC}}

For the quantitative discussion of Eqs.~(\ref{equ:eta2}) and (\ref{equ:zeta3}), 
one also has to specify the self-energy $\Pi (T)$. We adopt the quasiparticle 
model~\cite{Peshier94,Peshier96,Peshier00,Peshier02,Bluhm05,Bluhm07-1} 
with $\Pi (T) = T^2 \, G^2(T)/2$ and use the 
effective coupling~\cite{Peshier94,Peshier96,Peshier00,Peshier02} 
\begin{equation}
\label{equ:G2QPM}
 G^2(T) = \frac{16\pi^2}{\left[11\ln\left(\frac{T-T_s}{T_c/\lambda}\right)^2\right]}
\end{equation}
inspired by the QCD 1-loop running coupling for purely gluonic 
matter. In fact, by construction, the effective coupling $G^2(T)$ 
reproduces $g^2(T)$ at large $T$, where a temperature dependence is introduced 
into $g^2$ by considering it at variable renormalization point $2\pi T c$ with 
$c\in [1/2,2]$, and as scale 
$T_c/\Lambda_{\overline{MS}}=1.14$ from~\cite{Kaczmarek04,Kaczmarek05} 
is used. This implies, in particular, that $G^2(T)$ vanishes logarithmically as 
$T\rightarrow\infty$. In Eq.~(\ref{equ:G2QPM}), $\lambda$ and $T_s$ are two model 
parameters. 

\begin{figure}[t]
%\vspace{-4mm}
\centering
\includegraphics[scale=0.37]{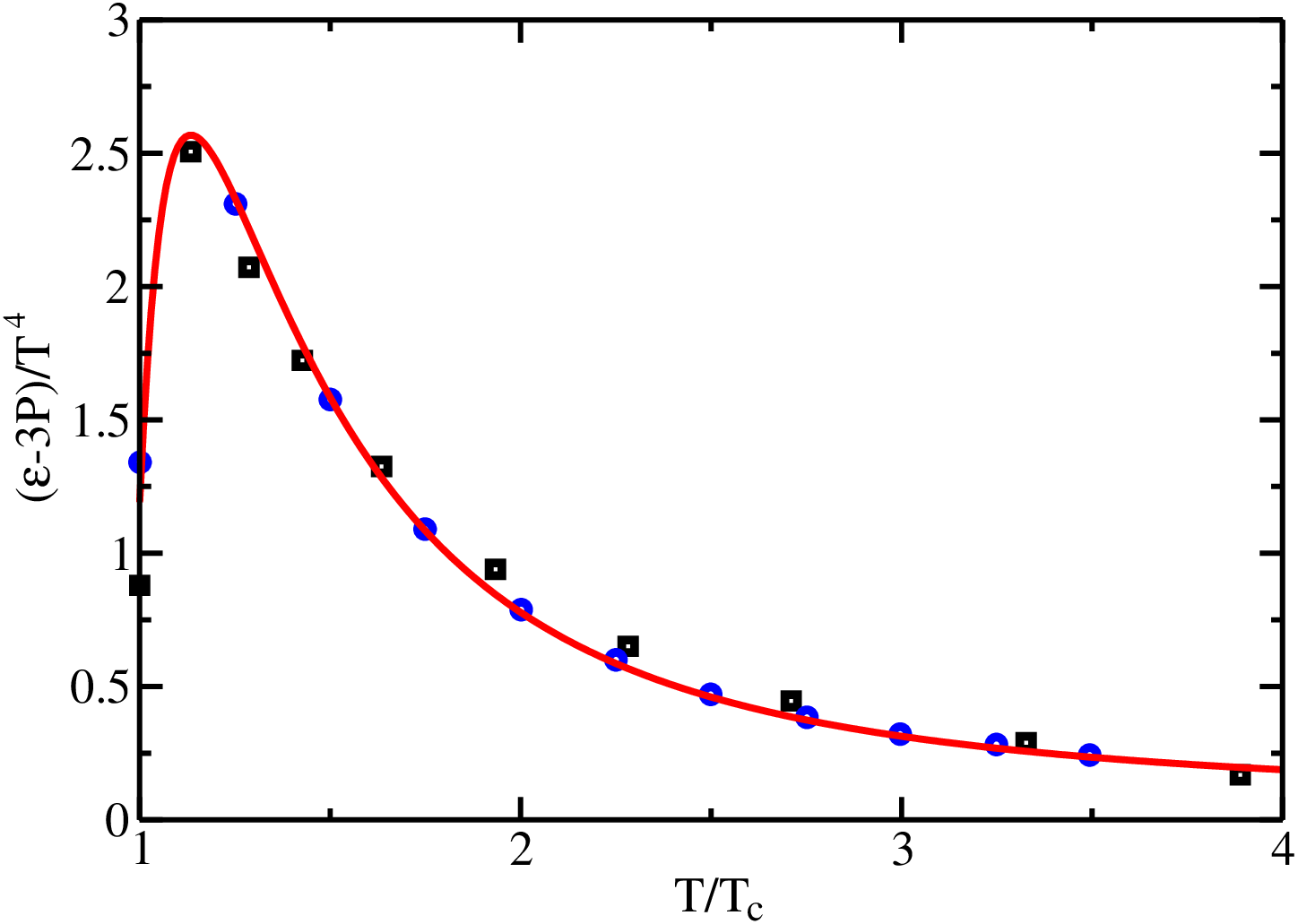} 
\caption[]{\label{fig:1}
(Color online) 
Comparison of QPM results (solid curve) for the scaled interaction 
measure $(\epsilon -3P)/T^4$ as a function of $T/T_c$ for the pure 
gluon plasma with $SU_c(3)$ lattice QCD results from~\cite{Boyd96} 
(boxes) and from~\cite{Okamoto99} (circles).}
\end{figure}
This quasiparticle model works impressively well in describing lattice QCD 
results for the equation of state and related quantities by adjusting $\lambda$ 
and $T_s$. Fig.~\ref{fig:1} 
exhibits the comparison of QPM results for the scaled interaction measure 
$(\epsilon -3P)/T^4$ as a function of $T/T_c$ with the corresponding 
lattice QCD results obtained in pure $SU_c(3)$ gauge 
theory~\cite{Boyd96,Okamoto99}. The model parameters were chosen as $\lambda=4.3$ 
and $T_s=0.73\,T_c$ with $T_c=271$ MeV~\cite{Karsch02}. In order to 
obtain $\epsilon$ and $P$~\cite{Bluhm07-1,Bluhm07-2}, we have evaluated 
\begin{eqnarray}
\label{equ:energy1}
 \epsilon(T) 
 & = & \int \frac{d^3 \vec{p}}{(2\pi)^3} E n(T) 
 + B(T) , \\ 
\label{equ:pressure1}
 P(T) 
 & = & \int \frac{d^3 \vec{p}}{(2\pi)^3 E} 
 \frac{\vec{p}^{\,2}}{3} n(T) 
 - B(T) , 
\end{eqnarray}
where $B(T)$ is the solution of
$\partial B(T) / \partial \Pi (T) = - \frac12
\int \frac{d^3 \vec p}{(2\pi)^3 \, E} n(T)$,
with integration constant $B(T_c) = 0.19\,T_c^4$.
Thereby, it is implicitly assumed that $B(T)$ depends on $T$ only via $\Pi (T)$. 
As evident from Fig.~\ref{fig:1}, both the maximum in 
the vicinity of $T_c$, which signals the strong deviation of $SU_c(3)$ Yang-Mills 
theory from the conformal limit $\epsilon=3P$~\cite{Megias09}, as well as the 
decline towards the conformal limit for asymptotically large $T$ observed 
in lattice QCD are well described within the QPM. The latter behaviour 
follows from the logarithmic decrease of the effective coupling $G^2(T)$ with $T$, 
while the position of the maximum is related to the inflection point of $p/T^4$ 
as a function of $\ln T$ and depends on the choices made for $\lambda$ and $T_s$. 

We note as an aside that the presence of dynamical quarks significantly influences 
the thermodynamics of the system, cf.~e.~g.~\cite{Borsanyi10} for corresponding 
lattice QCD calculations using physical quark masses. Similarly, it is expected 
that also the transport coefficients are considerably different for QCD compared 
to the pure gluon plasma. It would be interesting to study how these differences 
eventually translate into the specific viscosity coefficients. 

The squared speed of sound as a function of $T/T_c$ following the QPM result 
for the interaction measure exhibited in Fig.~\ref{fig:1} can be parametrized 
in the temperature interval $T_c<T\leq 4\,T_c$ by 
\begin{equation}
 \frac{\partial P}{\partial\epsilon} = 
 \frac13 \tanh\left[\sum_{n=1}^5 c_n(T/T_c-1)^n\right] \,,
\end{equation}
where the parameters read $c_1=4.78$, $c_2=-6.39$, $c_3=4.60$, 
$c_4=-1.57$ and $c_5=0.20$. The quality of this fit is quantified by 
$\chi^2/$d.o.f.$\,\,= 3.89\times 10^{-4}/16$. The function 
$\partial P/\partial\epsilon$ is monotonously rising with $T$, from 
small values just above $T_c$ towards $1/3$; at $T=4\,T_c$ the squared 
speed of sound is only $2.6\,\%$ smaller than $1/3$. At $T\rightarrow T_c^+$, 
it decreases fastly to zero as expected for a first-order phase 
transition~\cite{Fodor09}. 

The high temperature expansion of $P(T)$ from Eq.~(\ref{equ:pressure1}) 
for the pure $SU_c(3)$ plasma 
\begin{equation}
\label{equ:PhighT}
 P(T) = a T^4 \left(1- b G^2(T) + \mathcal{O}(G^4)\right) 
\end{equation}
with $a=d\pi^2/90$ and $b=15/(d\pi^2)$ was shown in~\cite{Peshier96} to be 
in agreement with the pQCD result~\cite{Arnold06}. 
The squared speed of sound at large $T$ can be evaluated from standard 
thermodynamic relations $\partial P/\partial\epsilon = 
dP/d\epsilon = (dP/dT)/(d\epsilon/dT) = (dP/dT)/(Td^2P/dT^2)$, 
where the individual derivatives read 
\begin{eqnarray}
\label{equ:pressdiff}
 \frac{dP}{dT} & = & 
 4aT^3(1-bG^2+\mathcal{O}(G^4)) - abT^4\frac{d G^2}{d T} + 
 \mathcal{O}(dG^4/dT) \,, \\
\label{equ:energydiff}
 T\frac{d^2 P}{dT^2} & = & 
 12aT^3(1-bG^2+\mathcal{O}(G^4)) -8abT^4 \frac{d G^2}{d T} 
 + \mathcal{O}(d^2G^2/dT^2,\,dG^4/dT) \,.
\end{eqnarray}
From Eq.~(\ref{equ:G2QPM}) and at large $T$ the following pattern is valid, 
$1 \gg G^2 \gg T(dG^2/dT) \gg T^2(d^2 G^2/d T^2)$. Thus, the squared speed of sound in 
the gluon plasma at large temperatures is given by 
\begin{equation}
\label{equ:soundvelo}
 \frac{\partial P}{\partial\epsilon} = \frac13 + 
 \frac{5}{36} b T \frac{dG^2}{dT} + \mathcal{O}(G^2TdG^2/dT) \,.
\end{equation}
The temperature derivative of the effective coupling 
follows from Eq.~(\ref{equ:G2QPM}) as 
\begin{equation}
 (T-T_s)\frac{d G^2}{d T} = -\frac{11}{8\pi^2} G^4(T) \,.
\end{equation}
Therefore, for large $T\gg T_s$ one finds 
$T(d G^2/dT) = - 11 G^4(T)/(8\pi^2) < 0$. In other words, 
$\partial P/\partial\epsilon\le 1/3$ is obtained 
from Eq.~(\ref{equ:soundvelo}). For asymptotically large $T$, 
$\partial P/\partial\epsilon$ approaches $1/3$.

\section{The shear viscosity \label{sec:eta}}

In~\cite{Arnold03}, the shear viscosity was obtained for the pure gluon 
plasma with $N_c=3$ colours at next-to-leading log order of small running 
coupling $g$ as 
\begin{equation}
\nonumber
 \eta_{NLL}=\frac{\eta_1T^3}{g^4\ln [\mu_*/(gT)]} 
\end{equation}
with $\eta_1=27.126$ and $\mu_*/T=2.765$. 
With $\tau$ from Eq.~(\ref{equ:tau}) and at large $T$, the expression for $\eta$ 
in Eq.~(\ref{equ:eta2}) exhibits parametrically the same dependencies on 
coupling and on temperature as the above perturbative result. 
Thus, within our formalism, by choosing the parameters in 
Eq.~(\ref{equ:tau}) appropriately, one 
can adjust the QPM to reproduce the above pQCD result well and then 
extrapolate the latter into the region close to $T_c$. 

Taking into account the behaviour of $G^2(T)$ for large $T$,
we choose $a_2=(\mu_*/T)^2$ in order to recover the constant 
in the logarithm of $\eta_{NLL}$. Furthermore, the QPM expression for the 
shear viscosity from Eq.~(\ref{equ:eta2}) can be written as 
$\eta=T^4 b(T)\, \tau$. Here, $b(T)$ is a dimensionless and 
with $T$ monotonically rising function. It approaches slowly its asymptotic 
limit $b(T\rightarrow\infty)=
d \Gamma(4)\zeta(4)/(30\pi^2)\simeq 1.404$ following the logarithmic decrease 
of $G^2(T)\rightarrow 0$ as $T\rightarrow\infty$. Thus, by setting 
$a_1\simeq 2.587\cdot 10^{-2}$, our $\eta\rightarrow\eta_{NLL}$ for 
$T\rightarrow\infty$. Variations in $a_1$, keeping $a_2$ fixed, allow to 
consider cases, where $\eta\rightarrow\eta_{NLL}$ for a large but finite 
temperature. 

\begin{figure}[t]
%\vspace{-4mm}
\centering
\includegraphics[scale=0.37]{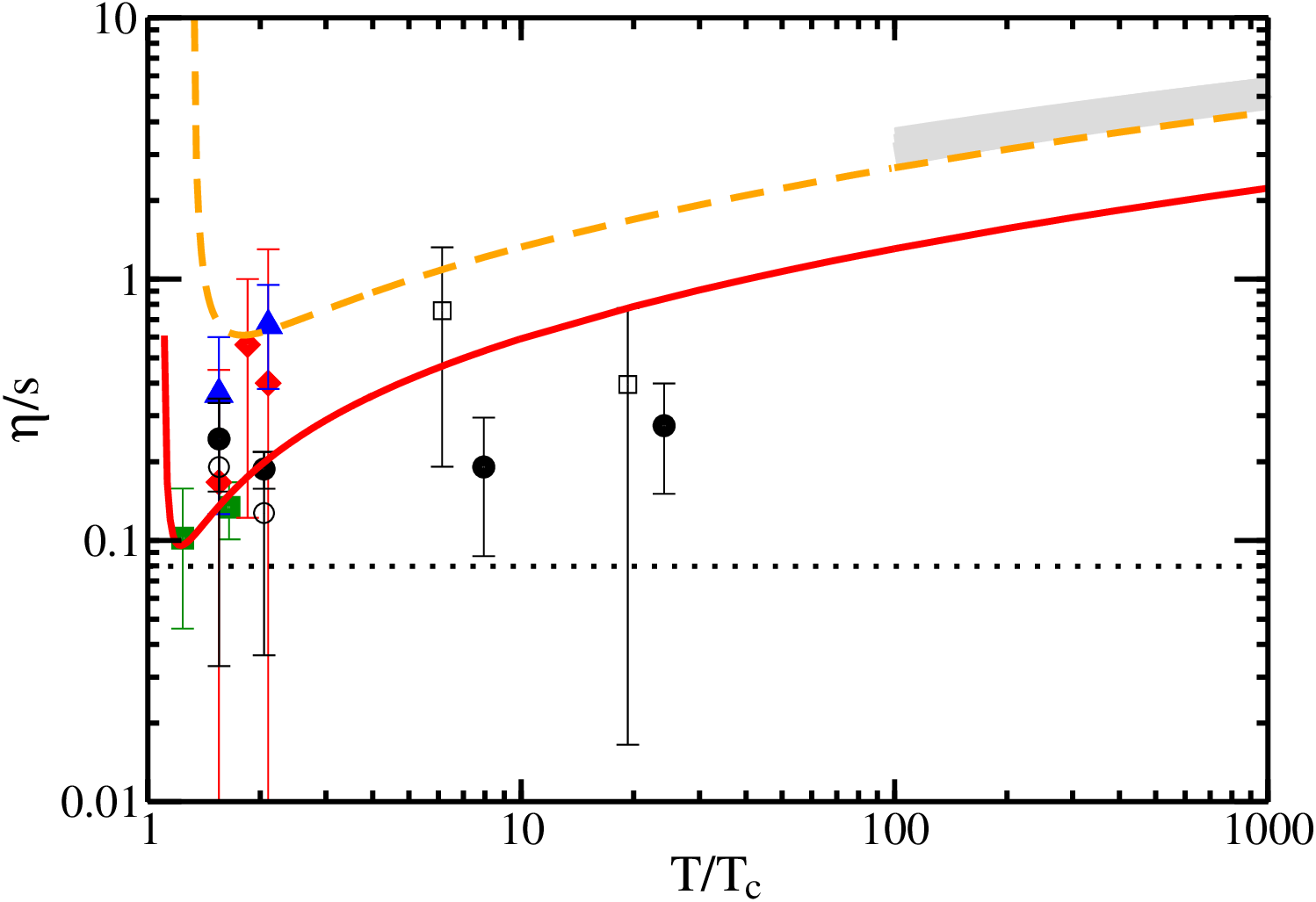} 
\caption[]{\label{fig:etadivs} 
(Color online)
Specific shear viscosity $\eta/s$ as a function of $T/T_c$. 
Dashed and solid curves exhibit the QPM results for Fit~1 and Fit~2, 
respectively (see text for details). The symbols denote lattice QCD 
data for pure $SU_c(3)$ gauge theory from~\cite{Meyer07} (full squares), 
from~\cite{Nakamura05} (diamonds and triangles) and from~\cite{Sakai07} 
(open squares, open and full circles). The dotted line exhibits 
the Kovtun-Son-Starinets bound~\cite{Policastro01,Kovtun03,Kovtun05} 
$(\eta/s)_{KSS}=1/(4\pi)$. The grey band at large $T$ indicates the 
perturbative QCD result $\eta_{NLL}/\tilde{s}$ with $\eta_{NLL}$ 
from~\cite{Arnold03} and entropy density $\tilde{s}$ from~\cite{Blaizot01}. 
The band was obtained by varying the renormalization point 
$\overline{\mu}$ in the QCD running coupling $g(\overline{\mu})$ in the 
range $\pi T \leq \overline{\mu} \leq 4\pi T$.} 
\end{figure}
The quantitative behaviour of the shear viscosity to entropy density ratio 
$\eta/s$ in the QPM as a function of $T/T_c$ is shown in Fig.~\ref{fig:etadivs} 
for different values of $a_1$ and $a_2$. The dashed curve corresponds to 
$a_1\simeq 2.587\cdot 10^{-2}$ and $a_2=(\mu_*/T)^2$ (Fit~1), while the 
solid curve corresponds to $a_1=3.85\cdot 10^{-2}$ and $a_2=2(\mu_*/T)^2$ (Fit~2). 
The entropy density $s$ in the QPM (cf.~\cite{Bluhm07-1,Bluhm07-2}) 
was calculated by using the parametrization (\ref{equ:G2QPM}) for $G^2(T)$, 
entering also the 
viscosities, such that results on equilibrium thermodynamics for 
the pure gluon plasma obtained in lattice QCD are reproduced (see 
Fig.~\ref{fig:1}). The model results are compared with 
perturbative QCD results and with recent lattice QCD data in 
Fig.~\ref{fig:etadivs}. 

As evident from Fig.~\ref{fig:etadivs}, the ratio $\eta/s$ is very sensitive to the 
values of $a_1$ and $a_2$ in Eq.~(\ref{equ:tau}). However, irrespective of 
the particular choices made for these parameters, the specific shear 
viscosity calculated from Eq.~(\ref{equ:eta2}) decreases with decreasing 
$T$ and exhibits a minimum at $T\gtrsim T_c$. The precise location 
of this minimum is determined by the value of $a_2$. For even smaller $T$, 
$\eta/s$ increases again monotonically as $T\rightarrow T_c^+$. This 
picture is qualitatively in agreement with model predictions for the confined 
phase reporting a monotonically decreasing $\eta/s$ for 
$T\rightarrow T_c^-$~\cite{Hostler09,Fernandez09-1,Chen07}. For both fits, the 
ratio $\eta/s$ is found to be significantly smaller than unity in a temperature 
region around its minimum, which is in contrast to naive extrapolations 
of pure pQCD results. 

With the parameters from Fit~1, the specific shear 
viscosity approaches for asymptotically large 
$T$ the result expected from perturbative calculations (see 
Fig.~\ref{fig:etadivs}). At smaller temperatures close to $T_c$, agreement 
with some but not the general bulk of available lattice QCD 
data~\cite{Nakamura05,Meyer07,Sakai07} is found. Exhibiting a minimum of 
$(\eta/s)_{min}\simeq 0.6$ at $T\simeq 1.75\,T_c$, $\eta/s$ rises to quite 
large values for $T\rightarrow T_c^+$. This, however, seems to be unfavoured 
by recent results obtained for $T\lesssim T_c$ in the confined 
phase~\cite{Hostler09,Fernandez09-1,Chen07}. 

Clearly, the extraction of transport coefficients from lattice QCD simulations 
is a difficult task~\cite{Moore08,Aarts02} and results seem far from being 
conclusive. Following the arguments in~\cite{Meyer08-2,Meyer09}, we focus on 
the lattice QCD results for $\eta/s$ from~\cite{Meyer07}. These results are 
best described with the parameters from Fit~2. In addition, as evident from 
Fig.~\ref{fig:etadivs}, this parametrization shows fairly nice agreement with 
available lattice QCD results over a wide range of temperatures (see 
also~\cite{Bluhm09}). 

\begin{figure}[t]
%\vspace{-4mm}
\centering
\includegraphics[scale=0.37]{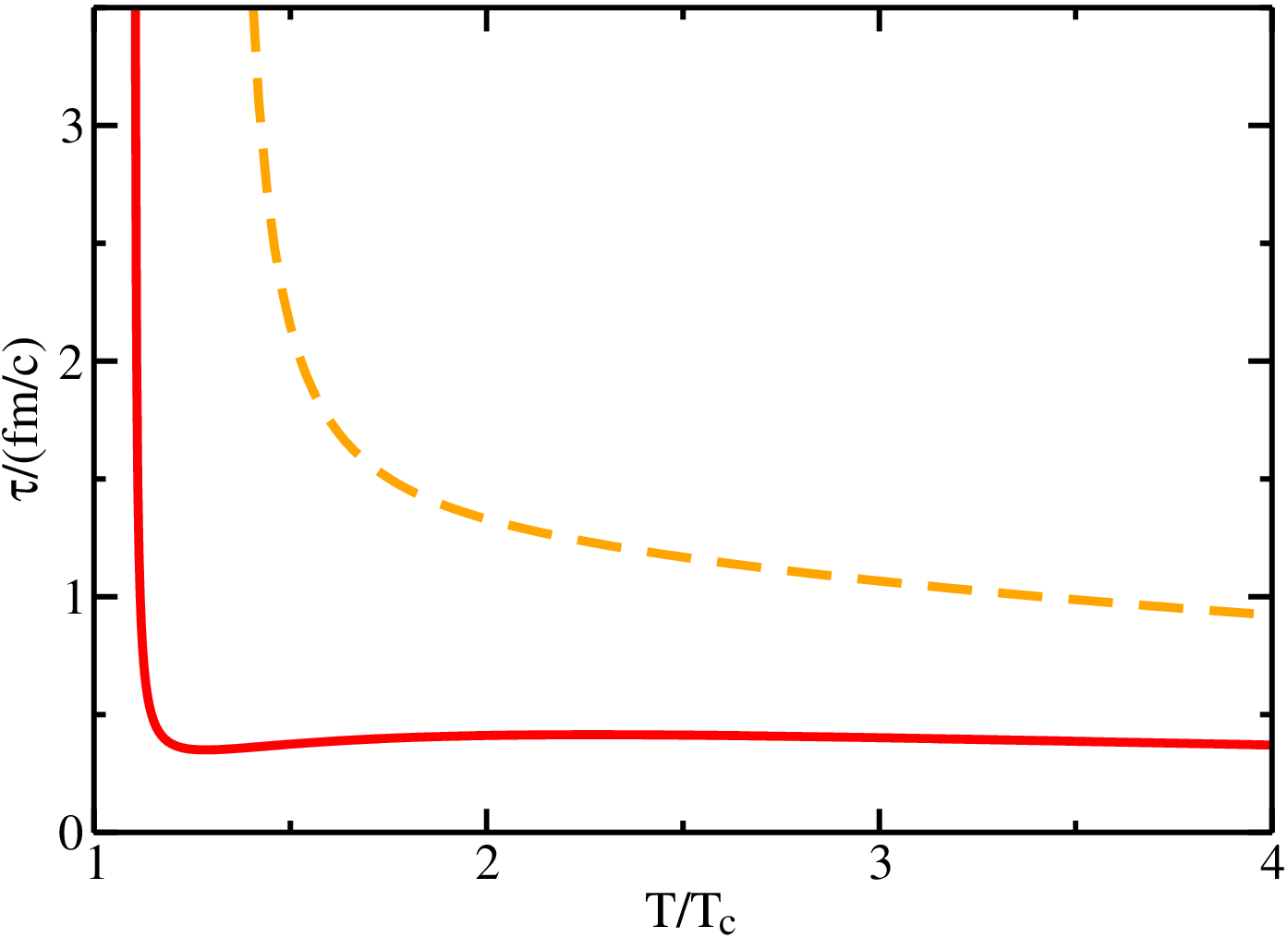} 
\caption[]{\label{fig:tau} 
(Color online)
Relaxation time $\tau$ from Eq.~(\ref{equ:tau}) for the parameters 
from Fit~1 (dashed curve) and Fit~2 (solid curve). See text for details.} 
\end{figure}
The behaviour of $\eta/s$ observed in Fig.~\ref{fig:etadivs} is a direct 
consequence of the particular form of $\tau$. In Fig.~\ref{fig:tau}, the 
relaxation time is exhibited as a function of $T/T_c$ for the 
different parameter sets corresponding to Fit~1 and Fit~2. 
While being rather flat at moderate 
temperatures above $T_c$, $\tau$ increases rapidly for decreasing temperatures. 
In particular in the case of Fit~2, a pronounced rise of $\tau$ in the vicinity 
of $T_c$ is observed. This behaviour is in line with the phenomenon of critical 
slowing down, indicating that the amount of time needed for relaxing energy and 
momentum density fluctuations increases rapidly near $T_c$. 

Interestingly, the sharp increase of $\tau$ close to $T_c$ is driven by the 
logarithm in Eq.~(\ref{equ:tau}). Without the logarithm, i.~e.~when 
assuming that $\tau$ depends solely on $(TG^4)^{-1}$, the relaxation time 
would simply decrease with decreasing $T$ in line with the increase 
of $G^2(T)$ for $T\rightarrow T_c^+$ according to Eq.~(\ref{equ:G2QPM}). 
However, close to $T_c$ the factor $\ln (a_2/G^2)$ becomes dominant and 
determines the behaviour of $\tau$. The precise temperature dependence is 
crucially driven by the parameter $a_2$, which can lead 
to potentially huge quantitative differences~\cite{Heinz94}. 
Increasing (decreasing) $a_2$ results in a shift of 
the sharp increase in $\tau$ to smaller (larger) $T$, apart from an 
overall decrease (increase) of $\tau$. 

As evident from Eq.~(\ref{equ:tau}), the relaxation time diverges at $T=T^*$ 
for which $G^2(T^*)\equiv a_2$. For the parametrization of $G^2(T)$ employed 
in Fig.~\ref{fig:1}, we find $T^*\simeq 1.32\,T_c$ (Fit~1) 
and $T^*\simeq 1.1\,T_c$ (Fit~2). For $T<T^*$, the relaxation time in 
Eq.~(\ref{equ:tau}) becomes negative, indicating that the used ansatz is not 
valid anymore. 

At large $T$, the behaviour of $\tau$ is dominated by its $(1/T)$ dependence 
such that $\tau\rightarrow 0$ for $T\rightarrow\infty$. However, in 
$\eta=T^4 b(T) \, \tau$, this temperature dependence of $\tau$ is compensated. 
The two different parametrizations Fit~1 and Fit~2 of $\tau$ result in a 
finite difference between the corresponding shear viscosities denoted by 
$\eta^{(1)}$ for Fit~1 (using $a_1^{(1)}$ and $a_2^{(1)}$) and $\eta^{(2)}$ 
for Fit~2 (using $a_1^{(2)}$ and $a_2^{(2)}$). At large $T$, the ratio 
$\eta^{(1)}/\eta^{(2)}$ reads 
\begin{equation}
\nonumber
 \frac{\eta^{(1)}}{\eta^{(2)}} = \frac{a_1^{(2)}}{a_1^{(1)}} 
 \left(1+\frac{\ln 2}{\ln [(\mu^*/T)^2/G^2]}\right) 
 \rightarrow \frac{a_1^{(2)}}{a_1^{(1)}} \simeq 1.49 \,.
\end{equation}
Therefore, by construction, the ratio $\eta^{(1)}/s$ approaches the pQCD result for 
asymptotically large $T$, whereas $\eta^{(2)}/s$ does not reach this limit at any $T$, 
remaining at least a factor $a_1^{(1)}/a_1^{(2)}$ smaller. 

With the parameters from Fit~2, we find for the relaxation time 
$\tau\simeq 0.4\,$fm$/$c for $T\geq 1.2\,T_c$, whereas it increases sharply 
beyond $5\,$fm$/$c as $T\rightarrow T_c^+$ (see Fig.~\ref{fig:tau}). 
Similar values were used in the hydrodynamic simulations~\cite{Song10} 
for discussing the influence of shear and bulk viscosities on observables 
measured in heavy-ion collisions such as the elliptic flow. Moreover, our 
results for $\tau$ are numerically comparable with results obtained 
in~\cite{Peshier04,Peshier05}, where a different ansatz for $\tau$ based 
on~\cite{Pisarski89,Lebedev90} was used reading 
\begin{equation}
\label{equ:tau2}
 \tau^{-1} = \frac{2N_c}{8\pi} T G^2 \ln \left(\frac{2c}{G^2}\right) \,.
\end{equation}
The parameter $c$ is determined by demanding $\tau\rightarrow\infty$ as 
$T\rightarrow T_c$, i.~e.~$c\equiv G^2(T=T_c)/2$. The results 
for $\eta/s$ from~\cite{Peshier05,Toneev10,Toneev10new}, where 
this ansatz for $\tau$ is employed, show a slightly flatter 
$T$ dependence compared with our Fit~2. However, Eq.~(\ref{equ:tau2}) 
differs from Eq.~(\ref{equ:tau}) in its parametric dependence on $G^2$. 
Therefore, as already discussed in~\cite{Peshier05}, the parametric 
dependence of $\eta$ on the coupling found in pQCD cannot be recovered 
with Eq.~(\ref{equ:tau2}) for $\tau^{-1}$. 

\begin{figure}[t]
%\vspace{-4mm}
\centering
\includegraphics[scale=0.37]{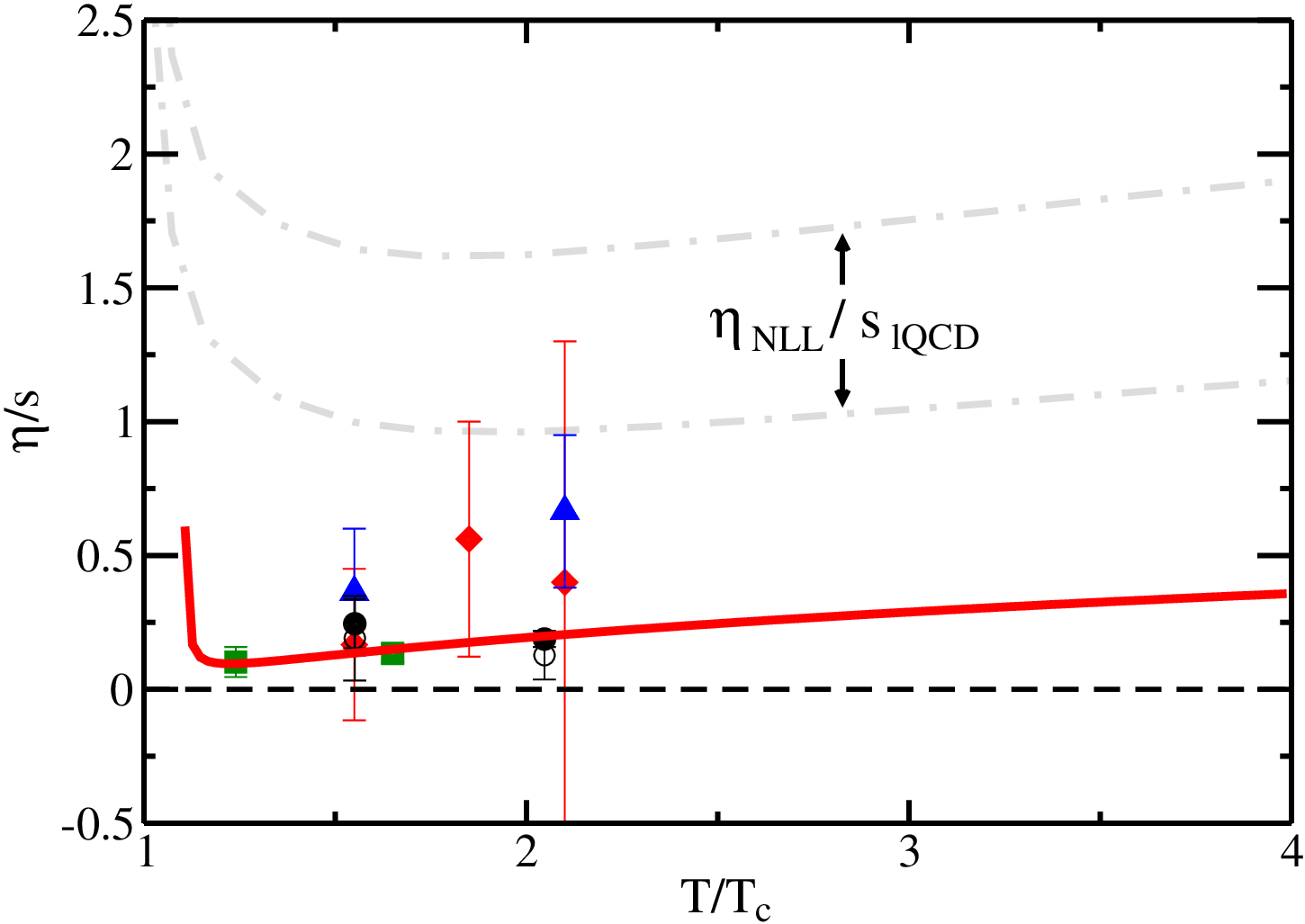} 
\caption[]{\label{fig:etadivsdetail} 
(Color online)
Specific shear viscosity $\eta/s$ as a function of $T/T_c$ as 
in Fig.~\ref{fig:etadivs} but zoomed into the temperature region 
near $T_c$. Symbols depict lattice QCD results as in Fig.~\ref{fig:etadivs} 
and the solid curve exhibits the QPM result for Fit~2. The area between 
the two grey dash-dotted curves shows the extrapolation of the 
pQCD result $\eta_{NLL}$ from~\cite{Arnold03} scaled by the entropy density 
from lattice QCD~\cite{Boyd96,Okamoto99}, where the renormalization point 
entering $\eta_{NLL}$ is varied between $\overline{\mu}=\pi T$ and 
$\overline{\mu}= 4\pi T$.} 
\end{figure}
As evident from Fig.~\ref{fig:etadivsdetail}, the ratio 
$\eta/s$ for Fit~2 increases very mildly with increasing $T$. 
For $T=1.5\,T_c$, we find $\eta/s=0.13$, which is a factor $1/3$ 
smaller than results obtained in~\cite{Gelman06} for 
the quark-gluon plasma with $N_f=3$ degenerate quark flavours. 
At $T=3\,T_c$, $\eta/s\simeq 0.29$ is still small and only a factor $3$ 
larger than the minimum value $(\eta/s)_{min}=0.096$. A similar value for 
the minimum was recently reported from a virial expansion 
approach~\cite{Mattiello09}. 

\section{The bulk viscosity \label{sec:zeta}} 
% last edited on 20/11/2010

The bulk viscosity exhibits different parametric dependencies on $T$ and 
on the coupling in comparison with the shear viscosity as observed in 
pQCD calculations~\cite{Arnold06,Arnold00,Arnold03}. In the QPM at large 
$T$, one finds with the expression for the squared speed of sound 
Eq.~(\ref{equ:soundvelo}) that Eq.~(\ref{equ:charge1}), which enters 
$\zeta$ quadratically, exhibits for thermal quasiparticle momenta 
the following structure 
\begin{equation}
\label{equ:charge2}
 \vec{p}^{\,2} \left( \frac{\partial P}{\partial\epsilon} 
 - \frac13 \right) - \frac12 \frac{\partial P}{\partial\epsilon} 
 T^4 \frac{\partial G^2}{\partial T^2} = \frac{5b}{36} T 
 \frac{\partial G^2}{\partial T} \vec{p}^{\,2} - \frac{1}{12} T^3 
 \frac{\partial G^2}{\partial T} 
 + \mathcal{O}(G^6) \,\sim\, T^2 G^4(T) \,.
\end{equation}
Together with $\tau^{-1}$ from Eq.~(\ref{equ:tau}), this implies for the 
bulk viscosity at large $T$, where $E \simeq |\vec{p}\,| \sim T$, that 
\begin{equation}
\label{equ:parambehav}
 \zeta \sim \frac{T^3 G^4(T)}{\ln(a_2/G^2(T))} \,.
\end{equation}
Keeping in mind the behaviour of $G^2(T)$ at large $T$, 
this result coincides with the leading log order parametric behaviour of 
$\zeta$ obtained in pQCD calculations by an expansion in 
inverse powers of $\ln(1/\alpha_s)$~\cite{Arnold06}. Nonetheless, our 
result for $\zeta$ exhibits the parametric dependencies on $T$ and $G^2(T)$ 
in Eq.~(\ref{equ:parambehav}) only, because it contains the subtracted mass 
term $(\Pi(T)-a)$ rather than the effective thermal mass $\Pi(T)$ of the 
quasiparticle excitations alone. 

\begin{figure}[t]
%\vspace{-4mm}
\centering
\includegraphics[scale=0.37]{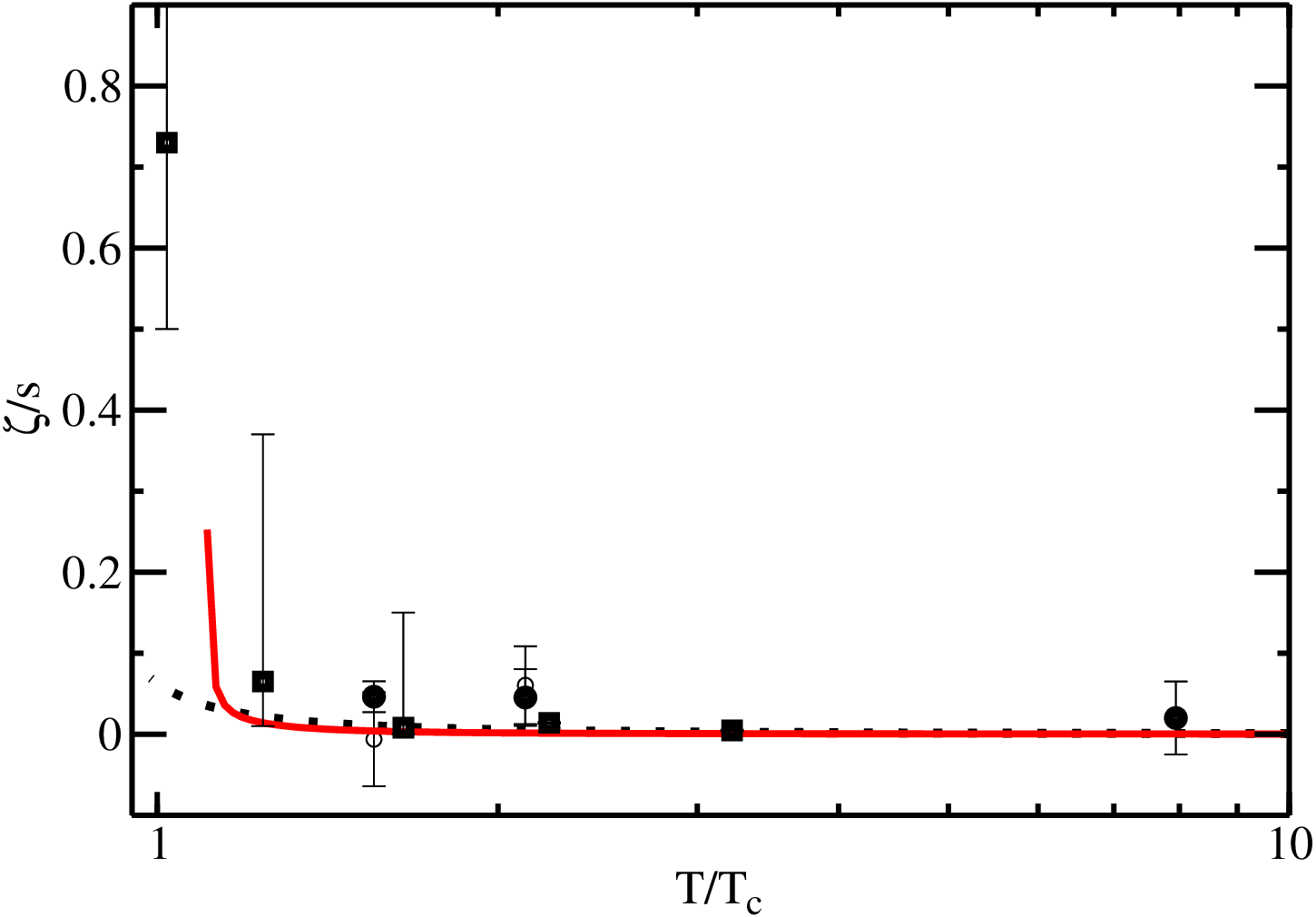} 
\caption[]{\label{fig:zeta}
(Color online) 
Specific bulk viscosity $\zeta/s$ as a function of $T/T_c$. The solid curve 
exhibits the QPM result for Fit~2. It is 
compared with available lattice QCD data from~\cite{Meyer08} (full squares) 
and~\cite{Sakai07} (open and full circles). The noticeable error bars for the 
results from~\cite{Meyer08} represent conservative upper and lower bounds 
given in~\cite{Meyer08}. The dotted curve shows for comparison results recently 
reported from holographic QCD~\cite{Guersoy09}.} 
\end{figure}
The QPM results for the specific bulk viscosity as a function of $T/T_c$ are 
depicted for Fit~2 in Fig.~\ref{fig:zeta} and compared with available lattice QCD 
results~\cite{Meyer08,Sakai07}. We find a positive but approximately vanishing ratio 
$\zeta/s$ for $T\geq 2\,T_c$ in line with lattice QCD. However, for $T\rightarrow 
T_c^+$, the specific bulk viscosity increases sharply following the behaviour of 
$\tau$. Similarly, for $T\rightarrow T_c^-$ in the confined phase, a monotonic 
increase of $\zeta/s$ was found in different model 
calculations~\cite{Hostler09,Fernandez09-2,Wiranata09,Chen09-1}. Both observations 
together suggest that for strongly interacting matter the ratio $\zeta/s$ develops 
a maximum in the vicinity of $T_c$~\cite{Kharzeev08,Karsch08,Romatschke09,Paech06}. 
Such a behaviour is, for instance, also seen in Lennard-Jones 
fluids~\cite{Meier05}. 

\begin{figure}[t]
%\vspace{-4mm}
\centering
\includegraphics[scale=0.37]{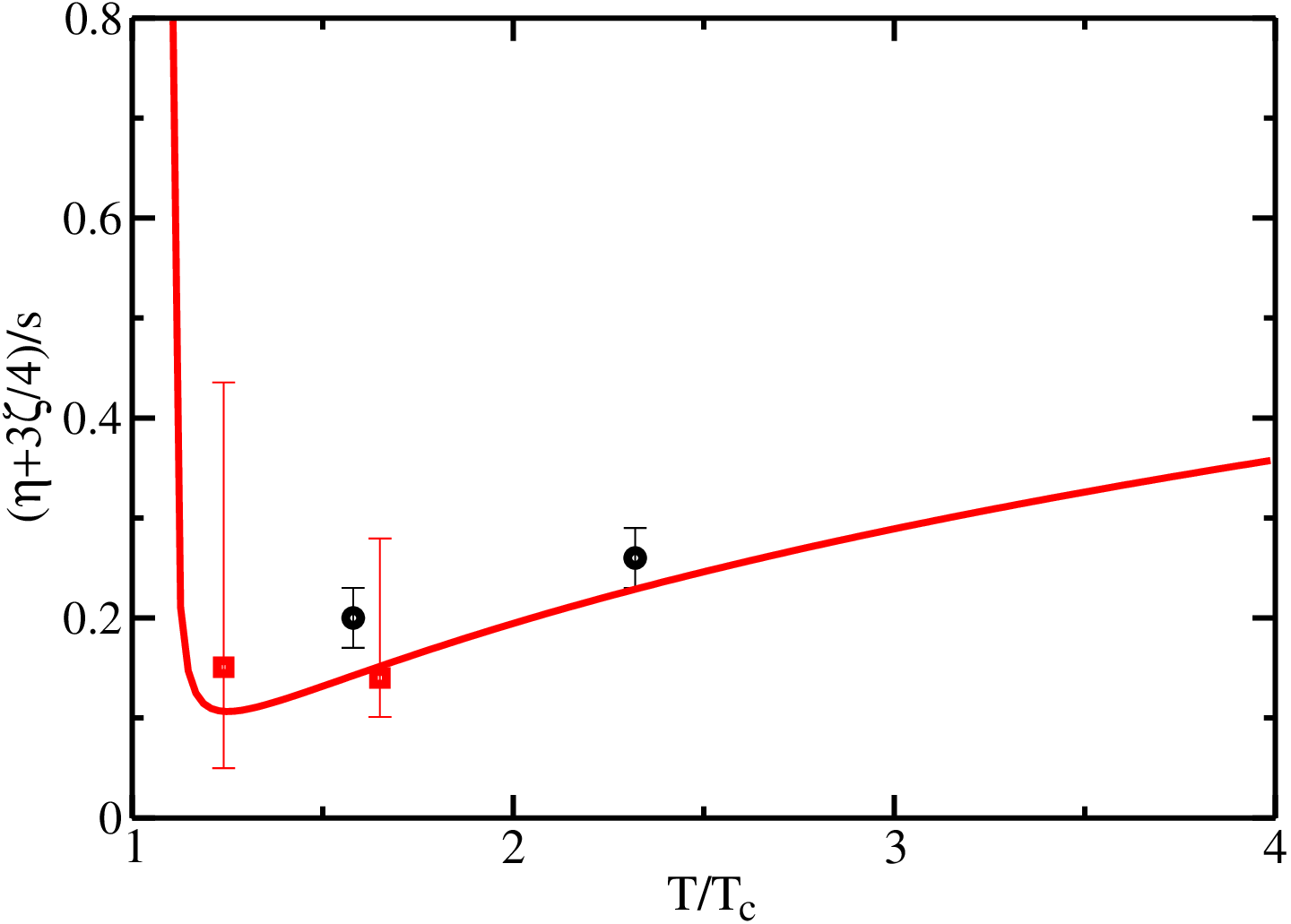} 
\caption[]{\label{fig:sound} 
(Color online)
Combined specific sound channel $(\eta+3\zeta/4)/s$ as a function of 
$T/T_c$. The QPM result for Fit~2 is shown by the solid curve. It is compared 
with available lattice QCD results taken either directly 
from~\cite{Meyer09} (circles) or from combining results of~\cite{Meyer07} 
and~\cite{Meyer08} (squares). The uncertainties in the latter arise from the 
upper and lower bounds in $\zeta/s$ depicted in Fig.~\ref{fig:zeta}.} 
\end{figure}

Even though the QPM results are in qualitative agreement with lattice QCD, 
nonetheless, for $T<2\,T_c$ they are systematically below the lattice QCD 
data~\cite{Meyer08}. However, the results from~\cite{Meyer08} are accompanied by 
noticeable uncertainties and a less rapid increase of $\zeta/s$ close to $T_c$ 
seems to be likely~\cite{Meyer09}. 
In fact, refined calculations~\cite{Meyer09} have been reported for the 
combined specific sound channel $(\eta+3\zeta/4)/s$, however, 
only for temperatures, where the contribution from $\eta$ is dominant. 
A comparison of QPM predictions for the specific sound channel with 
available lattice QCD data from~\cite{Meyer09} and from a combination 
of data from~\cite{Meyer07,Meyer08} is exhibited in Fig.~\ref{fig:sound}, 
showing fairly nice agreement. 

Our results for $\zeta/s$ and $\eta/s$ exhibit both a pronounced behaviour 
close to $T_c$, which might result in a sizeable influence of 
these transport coefficients on some observables measured in heavy-ion collisions. 
In fact, the observed possible rapid increase of $\eta/s$ with decreasing $T$ 
close to $T_c$ as well as the behaviour with $T$ in the confined 
phase~\cite{Hostler09,Fernandez09-1,Chen07} call for refinements 
in viscous hydrodynamic simulations~\cite{Demir09,Denicol10,Bozek10}. The smallness 
of $\zeta/s$ over a wide range of $T$, in contrast, could suggest that the 
impact of bulk viscous effects is less important. 
Indeed, in~\cite{Song10} a small influence of $\zeta/s$ on the shear viscous 
suppression of the elliptic flow was found. 
Although $\zeta/s$ becomes larger close to $T_c$, it increases less than 
the corresponding relaxation time (see Figs.~\ref{fig:tau} and~\ref{fig:zeta}) 
due to the thermodynamic integrals, cf.~Eq.~(\ref{equ:zeta3}). In other words, 
critical slowing down hampers the influence of bulk viscosity as long as the 
corresponding (negative) bulk viscous pressure is small enough to ensure the 
stability conditions of viscous hydrodynamics~\cite{Song10}. 
This effect also tempers the tendency of the fluid to become mechanically 
unstable against cavitation and clustering for large values of 
$\zeta$~\cite{Torrieri08-1,Torrieri08-2,Fries08,Rajagopal10}. 

\section{Conclusions \label{sec:4}} 
% last edited on 17/11/2010

Shear and bulk viscosities give important information about the transport 
properties of a medium. In this work, we calculated these transport coefficients 
for the pure gluon plasma within a quasiparticle model by assuming the plasma 
to be describable in terms of (quasi)gluon excitations with temperature dependent 
self-energy. 

The approach is based on an effective Boltzmann-Vlasov type kinetic theory for 
quasiparticle excitations with medium-dependent dispersion relation, which is 
consistent with the approach proposed in~\cite{Jeon95,Jeon96}. 
In local thermal equilibrium, this picture reduces to the quasiparticle model, 
which was employed successfully to describe lattice QCD equilibrium thermodynamics 
in~\cite{Bluhm05,Bluhm07-2,Bluhm08-1,Bluhm08-2}. Thermodynamic self-consistency 
in this model is a direct consequence of the consistency condition that must be 
imposed in order to fulfill energy and momentum conservation in the kinetic 
theory~\cite{Chakraborty10,Bluhm09}. 
We used expressions for shear and bulk viscosities derived within relaxation time 
approximation to the kinetic theory. 
In the shear viscosity, medium effects appear 
only implicitly via the dispersion relation. In the bulk 
viscosity, medium effects are in addition explicitly reflected by the subtracted 
mass term, which contains the derivative of the effective coupling with 
respect to the temperature. 

Including only contributions from elastic gluon-gluon scatterings to 
the relaxation time, we have shown that at large $T$ our results for the 
transport coefficients exhibit the same parametric dependencies on temperature 
and coupling as found in pQCD calculations~\cite{Arnold00,Arnold03,Arnold06}. 
For this to hold true, it is crucial that the bulk viscosity contains the 
subtracted mass term rather than the self-energy alone. 

In the temperature region close to $T_c$, fairly nice quantitative agreement 
with available lattice QCD results for the transport coefficients is found. 
The specific shear viscosity $\eta/s$ exhibits a minimum in the vicinity of 
$T_c$, while the specific bulk viscosity $\zeta/s$, being large close to $T_c$, 
falls off rapidly to its conformal limit for larger $T$. 

For the specific shear viscosity, we find a rather mild increase with 
temperature. In fact, the ratio $\eta/s$ is smaller than unity even at temperatures 
being a few times larger than $T_c$, e.~g.~at $T\simeq 3\,T_c$ it 
is still about $0.3$. Even though our investigations were limited to the case 
of a pure gluon plasma, one might expect that for thermal conditions relevant 
in heavy-ion collisions at the LHC (Large Hadron Collider), the ratio $\eta/s$ 
is small such that predictions for the LHC heavy-ion programme based on ideal 
hydrodynamic simulations are meaningful~\cite{Bluhm07-2,LHCpred,Aamodt10}. However, to 
make definite predictions it would be of importance to include also quark 
degrees of freedom into our quasiparticle approach. 

\subsection*{Acknowledgements}
% last edited on 08/11/2010

We gratefully acknowledge valuable and insightful discussions with 
P. Braun-Munzinger, B. Friman, S.~Jeon, H.~B.~Meyer, S.~Peign\'{e}, 
A.~Peshier, C.~Sasaki and U.~Wiedemann. We also thank J. Kapusta 
for expressing his opinion on the present work. 
The work is supported by BMBF 06 DR 9059, GSI-FE, the European Network 
I3-HP2 Toric and the Polish Ministry of Science.

\end{document}